\documentclass{ws-ijmpe}
\usepackage[super,compress]{cite}
\begin{document}

\title{PARITY VIOLATION AND ELECTRIC DIPOLE MOMENTS IN ATOMS AND MOLECULES}

\author{V. A. DZUBA}

\address{School of Physics, University of New South Wales\\
 Sydney 2052, Australia\\
V.Dzuba@unsw.edu.au}

\author{V. V. FLAMBAUM}

\address{School of Physics, University of New South Wales\\
 Sydney 2052, Australia\\
V.Flambaum@unsw.edu.au}

\maketitle

\begin{history}
\received{Day Month Year}
\revised{Day Month Year}
\end{history}

\begin{abstract}
We review the current status of the study of parity and time invariance violation 
in atoms, nuclei and molecules. We focus on parity
non-conservation in cesium and three of the most
promising areas of research: (i) parity non-conservation in a chain of
isotopes, (ii) search for nuclear anapole moments, and (iii) search
for permanent electric dipole moments (EDMs) of atoms and molecules,
which in turn are caused by either an electron EDM or nuclear $T,P$-odd moments
such as a nuclear EDM or nuclear Schiff moment. 
\end{abstract}


\ccode{PACS numbers:11.30.Er; 12.15.Ji; 31.15.A-}

\section{Introduction}

The study of parity and time invariance violation at low-energy in atoms,
molecules and nuclei is a relatively inexpensive
alternative to high-energy searches for new physics beyond the standard
model (see, e.g.  reviews~\refcite{Ginges,DK}).
Accurate measurements of the parity non-conservation (PNC) in atoms
is one of the most promising ways of exploring this path. It 
culminated in a very precise measurement of the PNC in the
cesium atom \cite{Wood}. There was even an indication that this
measurement showed some disagreement with the standard model and might
indeed lead to new physics\cite{BW}. It took considerable efforts
of several groups of theorists to improve the interpretation of
the measurement and resolve the disagreement in
favor of the standard model. The disagreement was removed when the
Breit\cite{Breit} and quantum electrodynamic corrections\cite{QED}
were included and the accuracy of the treatment of atomic correlations
was improved\cite{DFG02,PBD09}. 

The latter work\cite{PBD09} claimed a very small theoretical
uncertainty of just 0.26\%. Combined with the PNC measurement
\cite{Wood} it leads to perfect agreement between the measured weak nuclear
charge of the $^{133}$Cs nucleus and the value predicted by the standard
model. It also puts strong constraints on new physics beyond it. 

The PNC amplitude calculated in Ref.~\refcite{PBD09} is about 0.9\% smaller
than that of earlier calculations~\cite{DFG02,Nov,ND,KPT01}. The
difference is significant given that all earlier works agree very well
with each other and theoretical uncertainty in them is estimated to be
0.4 - 0.5\%~\cite{BW,DFG02}. The effect of this difference on the
agreement with the standard model is not large, however, there are
important implications for the constraints on new physics beyond it. 
The difference was attributed in Ref.~\refcite{PBD09} to the role of
high-order correlations. However, recently
in Ref.~\refcite{DBFR12} it was  found that the authors 
of Ref.~\refcite{PBD09} overlooked some important correlation
contributions to the PNC amplitude. The uncertainty of 0.26\% claimed
in Ref.~\refcite{PBD09} is based mostly on analysis of the dominating
contributions. The uncertainty due to the correlation corrections in other terms has been
underestimated.  The result of Ref.~\refcite{PBD09} when corrected as
suggested in Ref.~\refcite{DBFR12} is in perfect agreement with the previous
calculations\cite{DFG02,Nov,ND,KPT01}.
PNC in Cs is discussed in more detail in the next section.

It is unlikely that any new measurements of the PNC in atoms can
compete with the cesium experiment in accuracy of the 
interpretation. There are conflicting tendencies in the accuracy of
the measurements and the calculations. On the one hand, it is easier to get highly
accurate measurements for heavy atomic systems where the PNC
effect is larger.
On the other hand, advanced computational
techniques favor light alkaline atoms. 
The best accuracy for both calculations and measurements have been
achieved for cesium atom: 0.35\% for the measurements\cite{Wood} and
0.4 - 0.5\% for the calculations\cite{DFG02,Nov,ND,KPT01,DBFR12}.
A higher accuracy of the interpretation is possible for the rubidium atom,
which is a lighter analog of cesium. The rubidium atom has been considered
so far for a different type of PNC effect: the nuclear spin-dependent
PNC in the ground-state hyperfine transition, which is mostly due to
the nuclear anapole moment \cite{RbAM} (see section 4). For the spin-dependent PNC
accurate calculations are also possible, but less important
than for measurements of the weak nuclear charge. It is unclear
whether the weak nuclear charge of rubidium can be measured to high accuracy.
A heavier analog of cesium, the francium atom, is considered for both types
of PNC measurements, the effect of weak nuclear charge and the anapole
moment\cite{RbAM,FrPNC}. The accuracy of calculations for francium
could be, at best, as good as for cesium, but may be slightly lower due to
the larger relativistic effects which increase the Breit,
quantum electrodynamics (QED) and neutron distribution 
("neutron skin") corrections. There are
suggestions to measure  PNC in $s-d$ transitions of
Ba$^+\,$ \cite{BaII,Ba+} and Ra$^+\,$ \cite{KVI,Ba+} ions which have an electronic
structure similar to that of the cesium atom. Here again the accuracy
of the calculations  is unlikely to outperform cesium due to larger
correlations in $d$-states (only $s$ and $p$ states are involved in
cesium).  Note that some improvement here may be achieved using
experimental values 
of the electromagnetic E1 $p-d$ amplitudes in the sum over intermediate $p$ states
in the calculation of the $s-d$ PNC amplitude  \cite{Ba+} .
 
However, there are areas of symmetry breaking in atomic systems
where we do not need 
very high accuracy of atomic calculations: (i) the PNC measurements for a
chain of isotopes; (ii) the measurements of nuclear anapole moments; and
(iii) the measurements of the P,T-odd permanent electric dipole moments
of atoms and molecules. Accurate numerical atomic
calculations are not needed for the 
interpretation of the PNC measurements for  a chain of isotopes since
the change of the atomic wave function is very small and may be
estimated analytically in the ratio of PNC effects for different
isotopes. The 
interpretations of the anapole and electric dipole moment measurements
do rely on atomic or molecular calculations. However, high accuracy here is
not as important as for the measurements of the weak nuclear charge
(where we search for a small deviation from the Standard model predictions).

Below we briefly review parity non-conservation in cesium, and each of
the three subjects listed above. 

\section{Parity non-conservation in cesium}

\begin{table}[pt]
\tbl{Correlated PNC amplitude in Cs [in $10^{-11}i(-Q_W/N)$
    a.u.] in different calculations. Breit, QED and neutron skin
    corrections are not included.\label{t:CsK}}
{\begin{tabular}{@{}ll@{}} \toprule
\multicolumn{1}{c}{Value} & 
\multicolumn{1}{c}{Source and method} \\ 
\hline
0.908(9)  & CP+PTSCI, Ref.~\refcite{Nov} \\
0.909(9)  & CC SD, Ref.~\refcite{ND} \\
0.905(9)  & MBPT with fitting, Ref.~\refcite{KPT01} \\
0.9078(45) & CP+PTSCI, Ref.~\refcite{DFG02} \\
0.8998(24) & CC SDvT, Ref.~\refcite{PBD09} \\ \botrule
\end{tabular}}
\end{table}

Parity non-conservation in cesium is currently the best low-energy
test of the electroweak theory. It is due to the high accuracy of the
measurements and its interpretation. 
The experimental value for the PNC amplitude is\cite{Wood}
\begin{equation}
Im( E_{\rm PNC})/\beta = 1.5935\,(56) \ {\rm mV/cm}.
\label{eq:Wood}
\end{equation}
where $\beta$ is the vector transition amplitude. The amplitude $ E_{\rm
  PNC}$ is proportional to the weak nuclear charge $Q_W$
\begin{equation}
   E_{\rm PNC} = K (-Q_W/N),
\label{eq:K}
\end{equation}
where $K$ is the electron structure factor which comes from atomic
calculations, $N$ is the number of neutrons ($N=78$ for $^{133}$Cs). The
main challenge in the calculation of the PNC amplitude is an accurate
treatment of the inter-electron correlations. There are also minor
contributions arising due to the effects of the Breit interaction, quantum
electrodynamics and neutron skin corrections. We first consider
the dominant term (the correlated PNC amplitude) and add minor corrections
later. 

Table \ref{t:CsK} shows the most accurate values of the correlated
amplitude found in different calculations. 
The abbreviation CP+PTSCI stands for the correlation
potential\cite{CPM} combined with the perturbation theory in screened
Coulomb interaction, CC SD stands for the coupled cluster with single and
double excitations, CC SDvT means the coupled cluster with single, double
and valence triple excitations.
One can see that all
results except those of Ref.~\refcite{PBD09} agree with each other within
0.4\%. The result of Ref.~\refcite{PBD09} is 0.9\% smaller than another
most accurate result of Ref.~\refcite{DFG02}. Table~\ref{t:all} shows the
reason for this difference. The authors of Ref.~\refcite{PBD09} used the
sum-over-states approach in which the PNC amplitude is expressed as a
sum over complete set of single-electron states, including states in
the core and states above the core. Contribution of each state is
calculated independently of others and the use of the same
approximation for all terms is practically impossible. A very
sophisticated coupled cluster method was used to calculate the main
term (contributions of the four lowest states above the
core). Significantly less accurate approximation was used for the tail
(contribution of the states above $9p$) and a very simple Hartree-Fock
approximation was used for the core. The uncertainty of just 10\% was
assigned for the sum of the core and tail contributions. However, as
it was shown in Ref.~\refcite{DBFR12} the inclusion of the core polarization
effect for the core contribution changes even the sign of the core
contribution brining the total change to about 200\%, far beyond
the declared 10\% uncertainty. The tail contribution is also
significantly larger when the Brueckner-type correlations are
included. When the core and the tail contributions are corrected as
suggested in Ref.~\refcite{DBFR12}, the result of Ref.~\refcite{PBD09}
comes into excellent agreement with the previous calculations (see
``Subtotal'' line in Table~\ref{t:all} and Table \ref{t:CsK}, the corresponding numbers are 0.9079 and 0.9078).

\begin{table}[pt]
\tbl{All significant contributions to the $E_{\rm PNC}$ [in
  $10^{-11}i(-Q_W/N)$ a.u.] for Cs.\label{t:all}}
{\begin{tabular}{@{}lccl@{}} \toprule
\multicolumn{1}{c}{Contribution} & 
\multicolumn{2}{c}{Value} & 
\multicolumn{1}{c}{Source} \\
 & \multicolumn{1}{c}{Ref.~\refcite{PBD09}} & \multicolumn{1}{c}{Ref.~\refcite{DBFR12}} & \\
\hline
Core ($n<6$)   &  -0.0020\,(2) & 0.0018\,(8)  & Ref.~\refcite{PBD09,DBFR12} \\
Main ($n=6-9$) &  \multicolumn{2}{c}{0.8823\,(17)} & Ref.~\refcite{PBD09} \\
Tail ($n>9$)   &  0.0195\,(20) & 0.0238\,(35) & Ref.~\refcite{PBD09,DBFR12} \\
Subtotal       &  0.8998\,(25) & 0.9079\,(40) & Ref.~\refcite{PBD09,DBFR12} \\
Breit          & \multicolumn{2}{c}{-0.0055\,(1)} & Ref.~\refcite{Breit} \\
QED            & \multicolumn{2}{c}{-0.0029\,(3)} & Ref.~\refcite{QED} \\
Neutron skin   & \multicolumn{2}{c}{-0.0018\,(5)} & Ref.~\refcite{Breit} \\
Total          & 0.8906(26) & 0.8977\,(40) & Ref.~\refcite{PBD09,DBFR12} \\
\botrule
\end{tabular}}
\end{table}

Table \ref{t:all} also lists all other significant contributions to
the PNC amplitude in cesium.   
The resulting PNC amplitude is
\begin{equation}
 E_{\rm PNC} = 0.8977\,(45) \times 10^{-11}i(-Q_W/N)\,.
\label{eq:final}
\end{equation}
Note that the uncertainty coming from the analysis of Ref.~\refcite{DBFR12}
is smaller ($\pm 0.004$, see Table \ref{t:all}) due to the small
uncertainty of the main term claimed in Ref.~\refcite{PBD09}. For a conservative estimate we 
use in (\ref{eq:final}) a slightly larger uncertainty of
Ref.~\refcite{DFG02}. The central
points of Ref.~\refcite{DFG02} and Ref.~\refcite{DBFR12} are
practically identical. 

To find the value of the weak nuclear charge of $^{133}$Cs from 
(\ref{eq:Wood}) one also needs the value for the  vector transition
amplitude $\beta$.
The most accurate value for $\beta$
comes from the analysis\cite{beta2} of the Bennett and Wieman
measurements\cite{beta1} 
\begin{equation}
  \beta = 26.957\,(51)\, a_B^3,
\label{eq:beta}
\end{equation}
where $a_B$ is the Bohr radius.
Comparing (\ref{eq:final}), (\ref{eq:Wood}) and (\ref{eq:beta}) leads to
\begin{equation}
  Q_W(^{133}{\rm Cs}) = -72.58\,(29)_{\rm expt}\,(36)_{\rm theory}\,.
\label{eq:qw}
\end{equation}
This value is in a reasonable agreement with the prediction of
the standard model, $Q_W^{\rm SM} = -73.23\,(2)$\cite{SM} 
(see also Refs. ~\refcite{Thomas,Rosner,Marciano}). 
If we add theoretical and experimental
errors in (\ref{eq:qw}) in quadrature, the Cs PNC result deviates from
the standard model value by 1.4\,$\sigma$:
\begin{equation}
 Q_W-Q_M^{\rm SM} \equiv \delta Q_W = 0.65\,(46).
\label{eq:dq}
\end{equation}
For small deviations from the Standard Model values we may relate this to the
deviation in $\sin^2\theta_W$ using the simple relationship
$\delta Q_W \approx -4Z\, \delta(\sin^2\theta_W)$ which gives  $\delta (\sin^2\theta_W )= -0.0030\,(21)$ and 
\begin{equation}
  \sin^2\theta_W = 0.2356\,(21) \,.
\label{eq:sint}
\end{equation}
This is 1.4$\,\sigma$ off the standard model value
$0.2386\,(1)$\cite{SM} at near zero momentum transfer.

The new physics originated through vacuum polarization of the gauge boson
propagators, and  is described by the weak isospin conserving $S$ and isospin
breaking $T$ parameters\cite{Rosner}
\begin{equation}
 Q_W-Q_M^{\rm SM} = -0.800\,S-0.007\,T.
\label{eq:ST}
\end{equation}
At the 1$\sigma$ level (\ref{eq:dq}) leads to $S=-0.81\,(58)$.

Finally, a positive $\Delta Q_W$ could also be indicative of an extra
$Z$ boson in the weak interaction\cite{Marciano}
\begin{equation}
 Q_W-Q_M^{\rm SM} \approx 0.4(2N+Z)(M_W/M_Z{_{\chi}})^2.
\label{eq:Zxi}
\end{equation}
Using (\ref{eq:dq}) leads to $M_{Z_{\chi}}>700$ GeV/c$^2$.

\section{Chain of  isotopes}

It is convenient in this section to present the values measured in
atomic PNC-experiments in a form similar to (\ref{eq:K})
\begin{equation}
 E_{PNC} = k_{PNC} Q_W,
\label{eq:Epnc}
\end{equation}
where $k_{PNC}$ is an electron structure factor which comes from atomic
calculations, and $Q_W$ is the weak nuclear charge. Very sophisticated
calculations are needed for accurate interpretation of the
measurements as has been discussed in the introduction and previous section.
An alternative approach was suggested in
Ref.~\refcite{DFK86}. If the same PNC effect is measured for at least two
different isotopes of the same atom than the ratio
\begin{equation}
\mathcal{R} =\frac{E'_{PNC}}{E_{PNC}} = \frac{Q'_W}{Q_W}
\label{eq:R}
\end{equation}
of the PNC signals for the two isotopes does not depend on the electron
structure factor. It was pointed out however in Ref.~\refcite{Fortson}
that possible constraints on the new physics coming from the isotope ratio
measurements are sensitive to the uncertainties in the neutron
distribution which are sufficiently large to be a strong limitation
factor on the value of such measurements. The problem was addressed in
Ref.~\refcite{DP02} and more recently in Ref.~\refcite{BDF09}. The authors
of Ref.~\refcite{DP02} argue that experimental data on neutron
distribution, such as, e.g. the data from the experiments with
antiprotonic atoms\cite{Trzcinska}, can be used to reduce the
uncertainty. In the  more general approach of Ref.~\refcite{BDF09} nuclear
calculations are used to demonstrate that the neutron distributions
are correlated for different isotopes. This leads to significant cancelations of
the relevant uncertainties in the ratio (\ref{eq:R}).

The parameter $\mathcal{F}$ of the sensitivity of the ratio
(\ref{eq:R}) to new physics can be presented in the form
\begin{equation}
\mathcal{F}= \frac{h_p}{h_0} = \left( \frac{\mathcal{R}}{\mathcal{R}_0}
  -1 \right) \frac{NN'}{Z\Delta N},
\label{eq:F}
\end{equation}
where $h_p$ is the new physics coupling to protons ($\Delta Q_{new} =
Zh_p+Nh_n$), $h_0$ comes from the SM, $\mathcal{R}_0$ is the ratio
(\ref{eq:R}) assuming that each isotope has the same proton and
neutron distribution (no neutron skin),
$N$ and $N'$ are the numbers of neutrons in two isotopes, $Z$ is the
number of protons and $\Delta N=N'-N$.
The constraints on new physics parameter $h_p$ are affected by the
experimental error $\delta \mathcal{R}_{exp}$ and uncertainties in
$\mathcal{R}_0$ due to insufficient knowledge of neutron
distributions. However, as is argued in Ref.~\refcite{BDF09}, these errors are
correlated and strongly cancel each other.  Indeed, the distribution of the core neutrons is nearly the same in different isotopes. Significant changes happen in a small number of valence nucleons only. Estimations of
Ref.~\refcite{BDF09} show that corresponding contribution to
$\delta\mathcal{F}$ is in the range $10^{-3} \div 10^{-2}$ which is
about an order of magnitude smaller than the uncorrelated one. 
In the end, the isotope-chain measurements may be more sensitive to new
physics than current parity-violating electron scattering
measurements\cite{PVES} (by a factor of 10 for such atoms as Cs, Ba
and Dy). 

Experiments on isotope chains are in progress at Berkeley for Dy and
Yb atoms\cite{DyPNC,YbPNC}, at TRIUMF for Fr atoms\cite{FrPNC},
at Los Alamos for Yb$^+$ ions\cite{Yb+PNC},  and at Groningen (KVI)
for Ra$^+$ ions\cite{KVI}. There is an interesting recent suggestion to
measure the PNC in metastable Xe and Hg\cite{XeHg}. Most of these
experiments consider also the measurements of the nuclear anapole moment.

\section{Anapole moment}

The notion of the anapole moment was introduced by
B. Ya. Zeldovich\cite{zeldovich1957}. 
Nuclear anapole moment (AM) is the magnetic $P$ and $C$-odd, $T$-even
nuclear moment caused by the $P$-odd 
weak nuclear forces. 
Interaction of electrons with AM magnetic field (which may be called the
PNC hyperfine interaction) dominates the
nuclear-spin-dependent contribution to the atomic or molecular PNC effect.

First calculations of the nuclear AM and proposals for
experimental measurements were presented in
Refs.~\refcite{FKh80}-\refcite{HHM89}.
 Corrections to the AM interaction
with electrons due to
finite nuclear size were considered in Ref.~\refcite{FH93}. The status of the nuclear AM many-body calculations is presented in  reviews ~\refcite{Ginges,DK}. 
The authors of Refs.~\refcite{SF78,FKh85} (see also
Ref.~\refcite{labzovsky1978}) note in particular 
that the nuclear-spin-dependent PNC effect is strongly enhanced in diatomic
molecules due to mixing of the close rotational
states of opposite parity including mixing of $\Lambda$ or $\Omega$ doublets. 
The PNC effects produced by the weak charge are not enhanced. Therefore
the AM effect dominates PNC in molecules.
This greatly simplifies the detection of AM in diatomic molecules
compared to atoms. In atoms the AM effect is 50 times
smaller than the weak charge effect; AM effect is separated as a small
difference of the PNC effects in different hyperfine transitions.
A review of the parity and time invariance violation in diatomic
molecules (including the AM effect) can be found in Ref.~\refcite{KL95}. 

The idea of the AM contribution enhancement may be explained as follows. 
After the averaging over electron wave function the effective operator
acting on the angular variables may contain three vectors:
the direction of molecular axis ${\bf N}$, the electron angular
momentum ${\bf J}$
and nuclear spin ${\bf I}$. Scalar products ${\bf NI}$ and 
${\bf NJ}$ are $T$-odd and  $P$-odd. Therefore, they
are produced by the $T,P$-odd interactions discussed in the next section.
$P$-odd, $T$-even operator $V_P$ must be proportional to ${\bf N [J \times
  I]}$. It contains nuclear spin ${\bf I}$, therefore, the 
weak charge does not
contribute. The nuclear AM is directed along the nuclear
spin ${\bf I}$, therefore, it contributes to $V_P$. The matrix elements of 
${\bf N}$ between molecular rotational states are well-known, they produce
rotational electric dipole transitions in polar molecules. Therefore,
$V_P$ (induced by the magnetic interaction of the nuclear AM with
molecular electrons) mixes close  rotational-hyperfine states of opposite
parity.  
The interval between these levels is five orders of magnitude smaller
than the interval between the opposite parity levels in atoms
(by the factor $m_e/M$ where $m_e$ and $M$ are the electron and
reduced molecular masses), therefore PNC effects are five orders
of magnitude larger. 
Further enhancement may be achieved by a reduction of the intervals by an
external magnetic field\cite{FKh85}.

The effect is further enhanced for heavy
molecules. It grows with the nuclear charge as $Z^2A^{2/3}R(Z\alpha)$,
where $R(Z\alpha)$ is the relativistic factor which grows from $R=1$ at
low $Z$ to $R \sim 10$ for $Z > 80$ and the factor $A^{2/3}$ comes from the nuclear anapole, $A$ is the nucleon number. Good candidates for the
measurements include the molecules and molecular ions with the projection of the electron angular momentum on the molecular axis $\Omega=\Lambda_z+\Sigma_z=1/2$, i.e. in
$\Sigma_{1/2}$ ($\Lambda_z=0$) or $\Pi_{1/2}$ ($\Lambda_z=1$) electronic ground
states \cite{SF78,FKh85}, for example,  YbF, BaF, HgF, PbF, LaO,
LuO, LaS, LuS, BiO, BiS, YbO+, PbO+, BaO+, HgO+, etc.
Molecular experiments are currently
in progress at Yale\cite{DeMille} and Groningen KVI\cite{Jangman}.
An interesting idea of studying AM contribution to the NMR spectra of
chiral molecules was discussed in Ref.~\refcite{Nahrwold}.

Interpretation of the AM measurements requires electron structure
calculations. A number of semiempirical and {\em ab initio}
calculations have been performed for diatomic molecules of experimental
interests in Refs.~
\refcite{MAM1}-\refcite{Isaev}.
So far the only nuclear AM which has been measured is the AM of the
$^{133}$Cs nucleus. It is done by comparing PNC
amplitudes between different hyperfine structure sublevels in the same
PNC experiment  where the Cs weak charge is 
measured \cite{Wood} (the method was proposed in \cite{FKh80}). Interpretation of the
measurements\cite{Murray} indicates some problems. 
For example, the value of Cs AM is inconsistent with the limit on the AM
of Tl\cite{Tlanapole}.

To resolve the inconsistencies and obtain valuable information about
P-odd nuclear forces it would be very important to measure
anapole moments for other nuclei. In particular, it is important to
measure AM for a nucleus with an unpaired neutron (Cs and
Tl have unpaired protons). AM of such nucleus depend on different
combination of the weak interaction constants providing important
cross-check. Good candidates for such measurements include odd isotopes of
Ra, Dy, Pb, Ba, La, Lu and Yb. The Ra atom has an extra advantage
because of a strong 
enhancement of the spin-dependent PNC effect in the $^1$S$_0$ -
$^3$D$_2$ transition due to proximity of the
opposite-parity state $^3$P$^o_1$ ($\Delta E=5$ cm$^{-1}$)\cite{Ra}.

Experimental work is in progress for Rb and Fr at
TRIUMF\cite{RbAM,FrPNC} and for Dy and Yb at
Berkeley\cite{DyPNC,YbPNC}. Measurements for Xe and Hg are planned at
university of Crete\cite{XeHg}.

Calculations of the nuclear spin-dependent PNC amplitudes (including
the effect of AM) for a number of atoms and ions of experimental
interest were reported in Refs.\refcite{Ra}-\refcite{DF-hfs}.

\section{Electric dipole moment}

Permanent electric dipole moment (EDM) of neutron, atom or molecule
would violate both $P$ and $T$ invariance. Under conditions of the
$CPT$-theorem this would also mean  $CP$-violation. The
Kobayashi-Maskawa mechanism of the standard model leads to extremely small values
of the EDMs of the particles. It is also too weak to explain the
matter-antimatter asymmetry of the Universe. On the other hand, most
of the popular extensions to the standard model predict much larger EDMs which
are within experimental reach.
The EDM of an atom or a molecule is mostly due to
either electron EDM and T,P-odd electron-nucleon interactions in
paramagnetic systems (with non-zero total electron angular momentum $J$) or to the
$T,P$-odd nuclear forces in diamagnetic systems 
($J=0$; nuclear-spin-dependent e-N interaction contributes here too).
The existence of the  $T,P$-odd nuclear forces leads to the $T,P$-odd
nuclear moments in the expansion of 
the nuclear potential in powers of the  distance $R$ from the center of
the nucleus. The lowest-order term in the expansion, the nuclear EDM,
is unobservable in neutral atoms due to the total screening of the
external electric field by atomic electrons. It might be possible
however to observe the nuclear EDM in ions (see below). The first
non-vanishing electrostatic  terms which survives the screening in neutral systems are the octupole moment and the  Schiff moment. The octupole moment does not produce EDM in diamagnetic atoms. Below we discuss the effects of the  nuclear
and electron EDM and the Schiff moment.

\subsection{Nuclear EDM}

It was widely believed that one needs neutral particles (e.g.,
neutron, neutral atom or molecule) to study EDMs. This is because the
EDM is expected to be very small and it would be very hard to see the
effect of its interaction with external electric field on the
background of the much stronger interaction with the electric
charge. On the other hand, the EDM of neutral systems is very much
suppressed by the effect of screening of an external electric field by
electrons (Schiff theorem - see section 5.3). 
The Schiff theorem may be violated by the relativistic effect (which
dominates in the case of the electron EDM), the hyperfine interaction and the nuclear finite
size effect \cite{Schiff}. For example, the lowest-order $T,P$-odd nuclear moment, the
nuclear EDM is practically unobservable in the neutral systems (except for a
small contribution due to the hyperfine interaction calculated in Ref. \refcite{PGF}). First observable $T,P$
-odd nuclear moment, the Schiff moment, is non-zero due to the finite nuclear
size.

It is important therefore to explore the possibility of studying EDMs of
charged particles (e.g. muons or atomic ions). There are realistic
suggestions of this kind in Refs.~
\refcite{Khriplovich}-\refcite{Baryshevsky}
based on the use of ion storage rings. There are also proposals to measure EDM
in molecular ions \cite{Cornell}.

The external electric field is not totally screened on the nucleus of
an ion. Its value is
\begin{equation}
  E_N = \frac{Z_i}{Z} E_0,
\label{eq:EN}
\end{equation}
where $E_0$ is the external electric field, $E_N$ is the electric field at the
nucleus, $Ze$ is the nuclear charge, $Z_ie$ is the charge of the ion, $e$ is
proton charge. 
The formula (\ref{eq:EN}) can be obtained in a very simple way. The
second Newton law for the ion and its nucleus in the electric filed reads
\begin{equation}
  M_i a_i = Z_i e E_0, \nonumber \\
  M_N a_N = Z e E_N, \nonumber
\label{eq:Mi}
\end{equation}
where $M_i$ is the ion's mass,  $a_i$ is its acceleration,
$M_N$ is nuclear mass ($M_N \approx M_i$), and $a_N$ is its
acceleration. Since
the ion and its nucleus move together, the accelerations must be the
same ($a_i=a_N$),
therefore
\begin{equation}
 E_N = \frac{Z_i}{Z}E_0\frac{M_N}{M_i} \approx  \frac{Z_i}{Z} E_0. 
\label{eq:E_N}
\end{equation}
Quantum mechanical  derivations of this formula can be found in
Refs.~\refcite{DFSS88}-\refcite{FK12a}.
Numerical calculations of the screened electric field inside an atomic
ion were performed in a number of our works (see, e.g. Ref.~\refcite{DFSS88}). 

The Hamiltonian of the nuclear EDM ($d_N$) interaction with
the electric field is given by
\begin{equation}
  \hat H_d = d_N E_N = d_N\frac{Z_i}{Z}E_0.
\label{eq:Hd}
\end{equation}
Screening is stronger for diatomic molecules where we have an
additional suppression factor in  eq. (\ref{eq:E_N}),
$M_N/M_i=M_1/(M_1+M_2)$, 
where  $M_1$ and $M_2$ are the masses of the first and second
nucleus. 
For the average electric field acting on the first nucleus we obtain
\begin{equation}
 E_{1N}=\frac{Z_i}{Z_1}\frac{M_1}{M_1+M_2} E_0. 
\label{eq:e1n}
\end{equation}
A Quantum mechanical  derivation of this formula can be found in
Ref.~\refcite{FK12a}. 
Note that  the screening factor here contains both nuclear masses. This
indicates that the nuclear motion can not be ignored and the screening
problem is more complicated than in atoms. For example, in a naive ionic
model of a neutral polar molecule $A^+ B^-$, both ions
$A^+$ and $ B^-$ should be located in the area of
zero (totally screened) electric field since their average acceleration is
zero. This could make $A^+$ and $B^-$ EDM
unobservable even if they are produced by the nuclear Schiff moment or
electron EDM. In a more realistic molecular calculations the Schiff moment
and electron EDM effects are not zero, however, they may be significantly
suppressed (in comparison with a naive estimate of ionic EDM
in  a very strong field of another ion) and  the results of the calculations
may be unstable.

 Note also that contrary to atomic ions the contribution of the Schiff moment
to the T,P-odd effects in heavy molecular ions exceeds the contribution of the nuclear EDM 
\cite{FK12a}.  

In the case of monochromatic external electric field its frequency can
be chosen to be in resonance with the 
atomic electron excitation energy. 
Then for the effective Hamiltonian we can have
\begin{equation}
  \hat H_d = d_N E_N(t) \gg d_NE_0.
\label{eq:Hdt}
\end{equation}

\subsection{Electron EDM}

Paramagnetic atoms and molecules which have an unpaired electron are
most sensitive to the electron EDM. The EDM of such systems can be
expressed in the form
\begin{equation}
  d = K d_e,
\label{eq:d}
\end{equation}
where $d$ is the EDM of an atom or molecule, $d_e$ is electron EDM,
and $K$ is electron structure factor which comes from atomic
calculations. The factor $K$ increases with nuclear charge $Z$ 
as $Z^3\,$ \cite{Sandars65} times large relativistic factor
$R(Z\alpha)\,$\cite{Flambaum76} which may exceed the value of 3 in heavy
atoms. A rough estimate of the enhancement factor 
in heavy atoms with external $s_{1/2}$ or $p_{1/2}$ electron is 
$K \sim 3Z^3\alpha^2R(Z\alpha) \sim 10^2 -
10^3$\cite{Sandars65,Flambaum76}. 

Several orders of magnitude larger $K$ 
$\sim 10^7 -10^{11}$ exist in molecules due to the
mixing of the close rotational levels of opposite parity (including 
$\Lambda$-doublets)\cite{SF78}. Following Sandars
this enhancement factor is usually presented as a ratio of a  very large
internal molecular field to the external electric field which polarizes the
molecule.

The best current limit on electron EDM comes from
the measurements of the thallium EDM\cite{TlEDM} and reads
\begin{equation}
  d_e=(6.9 \pm 7.4) \times 10^{-28} e \ {\rm cm}.
\label{eq:deTl}
\end{equation}
Here the value $K=-585$\cite{Liu} was used for the interpretation of
the measurements. The value of $K$ for Tl is very sensitive to the
inter-electron correlations but three most complete
calculations\cite{Liu,DF09,PSK12} give very close results (there are
also calculations of Ref.~\refcite{Sahoo} which give smaller result). 

In contrast to paramagnetic atoms the diamagnetic (closed shell) atoms
are much less sensitive to electron EDM. This is because the only
possible direction of the atomic EDM in this case is along nuclear
spin and hyperfine structure interaction must be involved to link
electron EDM to nuclear spin. For example, for the mercury atom $K \sim
10^{-2}$\cite{FK85,MP87}.  However, due to very strong constraints on
the mercury EDM~\cite{HgEDM} the limit on electron EDM extracted from
these measurements is competitive to the Tl result (\ref{eq:deTl})\cite{HgEDM}
\begin{equation}
  |d_e| <  3 \times 10^{-27} e \ {\rm cm}.
\label{eq:deHg}
\end{equation}

New strong limits on the electron EDM were recently found from the
measurements for the YbF molecule\cite{HindsNature}.
Experiments are in progress to measure the electron EDM in
Cs\cite{Gould}, Fr\cite{FrEDM},  ThO\cite{ThO},
PbO\cite{PbO}  and
in solid-state experiments\cite{Sushkov}.

Molecular EDM due to electron EDM were calculated for a number of
diatomic molecules in Refs.\refcite{Meyer08}-\refcite{Petrov}.  

\subsection{Schiff moment}

Schiff moment is the lowest-order $T,P$-odd nuclear moment which appears
in the expansion of the nuclear potential when screening of the
 electric field by atomic electrons is taken into account.
This potential can be written as (see different  derivations, e.g. in
Ref.~\refcite{SAF,SAF08,FK12a,DK}) 
\begin{equation}
\phi (\mathbf{R}) = \int \frac{e\rho(r)}{|\mathbf{R}-\mathbf{r}|}d^3r
+ \frac{1}{Z}\left(\mathbf{d}\cdot\mathbf{\nabla}\right) 
\int \frac{\rho(r)}{|\mathbf{R}-\mathbf{r}|}d^3r,
\label{eq:phi}
\end{equation}
where $\rho(\mathbf{r})$ is the nuclear charge density normalized to $Z$,
and $\mathbf{d}$ is the
nuclear EDM. The second term in (\ref{eq:phi}) is the  screening. 
Taking into account the finite nuclear size the
dipole  term in the multipole expansion of (\ref{eq:phi}) can be written as\cite{FG02}
\begin{equation}
\psi(\mathbf{R}) = - \frac{3\mathbf{S}\cdot\mathbf{R}}{B} \rho(R),
\label{eq:psiS}
\end{equation}
where $B=\int \rho(R)R^4dr$ and
\begin{equation}
  \mathbf{S} = \frac{e}{10}\left[ \langle r^2\mathbf{r}\rangle -\frac {5}{3Z}
    \langle r^2 \rangle \langle \mathbf{r} \rangle \right]
\label{eq:S}
\end{equation}
is the Schiff moment. The expression (\ref{eq:psiS}) has no singularities
and can be used in relativistic calculations. 
More accurate expressions, which include corrections for the finite
nuclear size  $\sim Z^2 \alpha^2$, were  obtained
in Refs.~\refcite{FG02,FK12a,FK12b}. The authors also considered a  partial screening of the  octupole moment \cite{FK12b}. 
The Schiff moment is
caused by the $T,P$-odd nuclear forces. The dominant mechanism is
believed to be the $T,P$-odd nucleon-nucleon interaction. Another important
contribution comes from the EDMs of protons and neutrons. 

Schiff moment is the dominant nuclear contribution to the EDM of diamagnetic
atoms and molecules. The best limit on the EDM of diamagnetic atoms
comes from the measurements of the EDM of mercury performed in
Seattle\cite{HgEDM} 
\begin{equation}
  |d(^{199}{\rm Hg})| < 3.1 \times 10^{-29} |e| {\rm cm}.
\label{eq:HgEDM}
\end{equation}
Interpretation of the measurements requires atomic and nuclear
calculations. Atomic calculations link the EDM of the atom to its
nuclear Schiff moment. Nuclear calculations relate Schiff moment to
the parameters of the $T,P$-odd nuclear interactions. Summary of
atomic\cite{DFGK02,DFG07,DFP09} and nuclear\cite{FKS84,FKS86,octSM}
calculations for diamagnetic atoms of experimental 
interest is presented in Table~\ref{t:EDM}. 
To compare the EDM of different atoms we present only the results of our
nuclear calculations which all were performed by the same
method. For Hg and Ra there are several recent nuclear many-body
calculations available (see references in the most recent
calculation\cite{Ban}). 

The dimensionless constant
$\eta$ characterizes the strength of the $P,T$-odd nucleon-nucleon
interaction (in units of the weak interaction Fermi constant $G$) which is to be determined from the EDM measurements. Using
(\ref{eq:HgEDM}) and the data from the Table one can get
\begin{equation}
  S(^{199}{\rm Hg}) = (-1.8 \pm 4.6 \pm 2.7) \times 10^{-13} e \ {\rm
    cm}
\label{eq:SHg}
\end{equation}
and for the $T,P$-odd neutron-proton interaction
\begin{equation}
  \eta_{np} = (1\pm 3 \pm 2) \times 10^{-5}.
\label{eq:eta}
\end{equation}

\begin{table}[pt]
\tbl{EDMs of diamagnetic atoms of experimental interest.\label{t:EDM}}
{\begin{tabular}{@{}rr ccl@{}} \toprule
\multicolumn{1}{c}{$Z$} & 
\multicolumn{1}{c}{Atom} & 
\multicolumn{1}{c}{$[S/(\ e \ {\rm fm}^3)]$} &
\multicolumn{1}{c}{$\eta \ e \ {\rm cm}$} &
\multicolumn{1}{c}{Experiment}  \\
&&\multicolumn{1}{c}{$\times 10^{-17}e \ {\rm cm}$} &
\multicolumn{1}{c}{$\times 10^{-25}$} & \\
\hline
 2 & $^3$He    & $8\times 10^{-5}$ & $5\times10^{-4}$ & \\
54 & $^{129}$Xe & 0.38 & 0.7 & Seattle\cite{Seattle}, Ann
Arbor\cite{AnnArbor} \\
&&&& Princeton\cite{Princeton} \\
70 & $^{171}$Yb & -1.9 & 3 & Bangalore\cite{Bangalore},
Kyoto\cite{Kyoto} \\
80 & $^{199}$Hg & -2.8 & 4 & Seattle\cite{HgEDM} \\
86 & $^{223}$Rn & 3.3 & 3300 & TRIUMF\cite{RnEDM} \\
88 & $^{225}$Ra & -8.2 & 2500 & Argonne\cite{Argonne}, KVI\cite{KVIb} \\
88 & $^{223}$Ra & -8.2 & 3400 &  \\  \botrule
\end{tabular}}
\end{table}

\subsection{Nuclear  enhancement.} 
In some nuclei with the quadrupole deformation there is a close opposite parity level with  
the same angular momentum. This leads to an  enhancement of nuclear EDM \cite{HH} and
Schiff moment \cite{FKS84} up to an order of magnitude. 
Significantly larger enhancement is possible in nuclei with the octupole deformation\cite{octSM}. This can be
explained in a very simple way. Nuclear deformation creates a collective
intrinsic Schiff moment in the rotating nuclear reference frame
\begin{equation}
  S_{intr} \approx eZR_N^3 \frac{9\beta_2\beta_3}{20\pi\sqrt{35}},
\label{eq:Sintr}
\end{equation}
where $R_N$ is the nuclear radius, $\beta_2 \approx 0.2$ is the parameter
of the quadrupole deformation, and $\beta_3 \approx 0.1$ is the parameter
of the octupole deformation. The intrinsic Schiff moment (\ref{eq:Sintr})
does not violate $T$ or $P$ invariance and, if no $T,P$-odd 
interaction is present, it averages to zero in the
laboratory reference frame due to the  nuclear rotation. However, when
$T,P$-odd interaction is included, it can mix close rotational states
of opposite parity (which are similar to $\Omega$- doublet in diatomic molecules). Small energy interval between these states leads
to a strong enhancement of the nuclear Schiff moment in the laboratory
reference frame
\begin{equation}
\label{eq:Sbig}
  S_{lab} \sim \frac{\langle + | H_{PT} | - \rangle}{E_+ - E_-}
  S_{intr} \sim \\
0.05e\beta_2\beta_3^2ZA^{2/3}\eta r_0^3\frac{\rm eV}{E_+ - E_-} \sim
700\times 10^{-8} \eta e \ {\rm fm^3}, \nonumber
\end{equation}
where $r_0=1.2$ fm is the inter-nucleon distance $|E_+ - E_-| \sim$ 50
keV. The estimate (\ref{eq:Sbig}) is about 500 times larger than the
Schiff moment of a spherical nucleus like Hg.

It was pointed out in Ref.~\refcite{softSM} that the octupole deformation
doesn't need to be static. Soft octupole vibrations lead to similar
enhancement.

 Large values of the Schiff moment for Ra and Rn (see
Table~\ref{t:EDM}) are due to the nuclear octupole deformation.
Some additional enhancement of the atomic  EDM is due to the large nuclear charge.
The nuclear enhancement may also manifest itself in molecules,
e.g. in RaO \cite{RaO}, where the $T,P$-odd nuclear spin- molecular axis interaction
exceeds that in the experimentally studied TlF \cite{TlF,DFGK02} about 500 times.

\section*{Acknowledgments} 
This work was supported by the Australian Research Council.

\end{document}